\documentclass{article}
\usepackage[preprint,position]{neurips_2026}

\usepackage[utf8]{inputenc}
\usepackage[T1]{fontenc}
\usepackage{hyperref}
\usepackage{url}
\usepackage{booktabs}
\usepackage{tabularx}
\usepackage{array}
\usepackage{makecell}
\usepackage{xcolor}
\usepackage{graphicx}
\usepackage{tikz}
\usetikzlibrary{shapes.geometric, arrows.meta, positioning, fit, backgrounds, calc}
\usepackage{microtype}
\usepackage{enumitem}
\setlist{leftmargin=1.5em, itemsep=2pt, topsep=3pt}

\definecolor{tier1}{HTML}{1E5E38}
\definecolor{tier2}{HTML}{7A4F00}
\definecolor{tier3}{HTML}{7A1A1A}
\definecolor{neuripsnavy}{HTML}{1A3A5C}
\definecolor{lightnavy}{HTML}{E8EFF7}
\definecolor{boxgray}{HTML}{F4F6F9}

\title{NeurIPS Should Require Reproducibility Standards\\[2pt]
for Frontier AI Safety Claims}

\author{%
  Varad Vishwarupe\thanks{Corresponding author: \texttt{varad.vishwarupe@cs.ox.ac.uk}} \quad Ivan Flechais \quad Marina Jirotka \quad Nigel Shadbolt \\
  Department of Computer Science, University of Oxford
}

\begin{document}
\maketitle

\begin{abstract}
Frontier AI safety claims -- published assertions that a highly capable
general-purpose model is below a threshold of concern, adequately
mitigated, or suitable for release -- increasingly shape model
deployment, governance, and public trust. Yet the artefacts needed to
evaluate them are routinely withheld, producing an \emph{evidential
inversion}: the most consequential claims in AI safety are often the
least reproducible. This position paper argues that NeurIPS should
require reproducibility standards for papers making such claims,
treating non-reproducibility not as a transparency preference but as
an evaluation-methodology failure. The 2026 International AI Safety
Report \citep{iaisr2026} concludes that reliable pre-deployment
safety testing has become harder to conduct and that models now
distinguish test from deployment contexts; the 2025 Foundation Model
Transparency Index \citep{wan2025fmti} reports a sector-average
transparency score of 40/100 with no major developer adequately
disclosing train--test overlap; contemporaneous measurement-theory
work shows that attack-success-rate comparisons across systems are
often founded on low-validity measurements
\citep{chouldechova2025asr}. We propose a three-tier disclosure
framework, distinguishing public, controlled, and claim-restricted
disclosure, paired with a mandatory claim inventory, scope statements,
and a phased implementation path with graduated sanctions. The
framework treats secrecy and openness as endpoints of a spectrum,
with controlled review (via a federated colloquium of qualified
secure-review hosts) covering claims whose artefacts cannot be
released publicly, and right-scaling claims whose artefacts cannot be
reviewed even confidentially. The standard the community applies to
its most consequential claims should be at least as high as the
standard it applies to its least.
\end{abstract}

% ============================================================
\section{Introduction}

\textbf{Position: NeurIPS should require reproducibility standards for
frontier AI safety claims.} When public release of evaluation artefacts
would materially increase capability, security, or misuse risk, NeurIPS
should not waive reproducibility altogether; it should require
controlled disclosure and narrow the permissible scope of claims
accordingly. This is not a call to force public release of dangerous
artefacts. It is a call to stop treating non-reproducibility as
epistemically costless when the claim being made is precisely that a
frontier model is safe enough, below a threshold of concern, or
adequately mitigated for deployment. The paper's scope is
methodological and review-time: it does not address governance of
frontier AI deployment, propose new evaluation methods, or redefine the
conference's remit.

The argument follows from a pattern visible in the 2025--2026 evidence
base. Twelve leading AI companies published or updated Frontier AI
Safety Frameworks in 2025 \citep{iaisr2026}. Each contains safety
claims; many are consequential and used to justify release decisions.
Yet across the same period, transparency on the methodological details
of those claims has \emph{declined}: the 2025 Foundation Model
Transparency Index \citep{wan2025fmti} reports an average sector score
of 40/100 (down 9 points year-over-year for weighted-average flagship
developers), six major developers (including OpenAI, Google, and xAI)
at zero on model-information disclosures, and no developer adequately
reporting train--test overlap. Independent measurement-theory work has
shown that attack-success-rate comparisons in red-teaming are often
founded on apples-to-oranges comparisons or low-validity measurements
\citep{chouldechova2025asr}, and that frontier models can ``fool all
output-based monitors used to detect jailbreaks''
\citep{panfilov2025strategic}. The result is what we call an
\emph{evidential inversion}: in most of ML, the more consequential the
empirical claim, the stronger the norm of disclosure and rerunnability;
in frontier AI safety, the relationship is often inverted.

The argument is timely for three reasons. First, frontier-AI governance
now depends heavily on published safety frameworks and external
evaluations, and the 2026 International AI Safety Report
\citep{iaisr2026} notes that frameworks vary substantially in scope and
that external assessments ``remain limited and non-standardised.''
Second, major policy instruments (the Seoul AI commitments
\citep{seoul2024}, the EU GPAI Code of Practice \citep{euaioffice2025},
the California Report on Frontier AI Policy
\citep{bommasani2025california}, IASEAI'26 \citep{iaseai2026}, and the
February~2026 Delhi Declaration \citep{delhi2026}) have begun to
formalise transparency duties for advanced models, and explicitly
treat peer-reviewed published research as input to governance. Third,
the NeurIPS 2026 Position Paper Track is explicitly broadening toward
technically grounded positions \citep{neurips2026track}, and the 2025
cohort included multiple exemplars that redefined evaluation,
measurement, and rigor norms for frontier AI
\citep{chouldechova2025asr,olteanu2025rigor,dolin2025postdep,
schaeffer2025refutations,bommasani2025consensus}. This paper sits in
direct continuity with that cohort.

\paragraph{Frontier vs.\ AGI}
We use \emph{frontier AI safety claims} rather than AGI safety claims
deliberately. ``Frontier'' is the operative vocabulary in current
providers' risk frameworks, governments' policy instruments
\citep{euaiact2024,bommasani2025california}, and the 2026 IAISR
\citep{iaisr2026}. ``AGI'' is operationally unstable; a NeurIPS
position paper must define a reviewable scientific object.

\paragraph{Contributions}
We contribute (1)~an operational definition of a frontier AI safety
claim with a quantitative anchor, a functional-frontier backstop, and
a normative/descriptive decision rule (Section~\ref{sec:def}); (2)~a
diagnosis of the evidential inversion grounded in 2025--2026 empirical
evidence (Section~\ref{sec:problem}); (3)~the case for NeurIPS as the right
site (Section~\ref{sec:why}); (4)~a three-tier disclosure framework with
claim inventory, scope statements, and a federated-colloquium model
for T2 review (Section~\ref{sec:proposal},
Appendix~\ref{app:federated}); (5)~a phased implementation path with
disclosure labels and graduated sanctions (Section~\ref{sec:impl}); and
(6)~responses to five strong objections (Section~\ref{sec:alt}).
Appendices~B--F provide an illustrative T2 submission, a precedent
comparison, a reviewer governance protocol, a comparison with
technical alternatives, and the federated review infrastructure.

% ============================================================
\section{Frontier AI Safety Claims: Definition and Taxonomy}
\label{sec:def}

A \emph{frontier AI safety claim} is a published assertion about the
safety, controllability, or societal-risk mitigation of a highly
capable general-purpose model or frontier deployment, where the
assertion is liable to inform public governance, standards, or
procurement decisions. The definition has four operative components.

\paragraph{Highly capable general-purpose model (quantitative anchor)}
We follow the EU AI Office's definition of a general-purpose AI model
as one trained on broad data at significant compute and capable of
performing a wide range of tasks \citep[Art.\ 3(63)]{euaiact2024}. To
reduce reviewer subjectivity, ``frontier'' is anchored to one of two
operative thresholds, either of which suffices: (a)~the model exceeds
the EU AI Act's systemic-risk training-compute threshold of
$10^{25}$~FLOPs \citep[Art.\ 51(2)]{euaiact2024} or California
SB-53's $10^{26}$~FLOPs threshold; or (b)~the paper presents evidence
of emergent dangerous capabilities (CBRN uplift, autonomous
cyber-offence, self-replication, or strategic-deception capability)
regardless of compute. Both anchors track the operative definitions
used by major governance frameworks
\citep{anderljung2023,iaisr2026}.

\paragraph{Safety, controllability, or societal-risk mitigation}
This covers harmlessness evaluations, alignment benchmark results,
red-teaming summaries, controllability assessments, and risk-card
conclusions. The defining feature is that the claim is
\emph{normative}: it asserts not merely that the model achieves $X$\%
on a benchmark (descriptive) but that the model is \emph{safe enough}
or \emph{adequately mitigated} (normative). To reduce ambiguity at
submission time, we apply an operative \emph{decision rule}: a claim
is normative (and in scope) if the paper's abstract, conclusion, or
policy implications argue for a deployment decision, threshold
crossing, or mitigation adequacy. Purely descriptive numbers (e.g.,
MMLU scores) that authors do not use to license such conclusions are
out of scope.

\paragraph{Liable to inform governance, standards, or procurement}
The operative test is what the claim is reasonably likely to be used
for, not what authors intend. The capability-based anchor in (b)
operates as a \emph{functional-frontier backstop}: a claim is also in
scope below the compute threshold if it concerns dangerous-capability
uplift, systemic-risk mitigation, release-justification, or
evaluation sufficiency for a model intended for broad deployment.
This closes the principal gaming pathway (training a small but
dangerously capable model under the FLOP gate) without requiring
reviewers to second-guess training compute. To prevent a converse
gaming pathway in which authors avoid the capability anchor by simply
not claiming emergence, the trigger is textual rather than
interpretive: a paper is in scope if it reports any evaluation of
CBRN uplift, autonomous cyber-offence, self-replication, or
strategic-deception capability, regardless of headline finding.

\paragraph{In-scope claim categories}
At minimum: dangerous-capability assessments; threshold-crossing or
threshold-absence claims; mitigation-adequacy claims;
evaluation-sufficiency claims used to support a release decision;
residual-risk claims; and absence-of-concerning-capability claims under
a stated evaluation regime
\citep{bengio2024extreme,vaccaro2026uplift,kolt2024}.

% Table 1: Taxonomy of Frontier AI Safety Claims
\begin{table}[ht]
\caption{Five principal types of frontier AI safety claims, the artefacts required to evaluate them, and the typical disclosure gap. Reproducibility failures are concentrated where claim consequence is highest.}
\label{tab:taxonomy}
\centering\small
\begin{tabularx}{\linewidth}{@{}p{2.5cm} X X X@{}}
\toprule
\textbf{Claim type} & \textbf{Example claim} & \textbf{Required artefacts} & \textbf{Typical disclosure gap} \\
\midrule
Harmlessness evaluation
& ``Our model refuses 94\% of harmful prompts on HarmBench''
& Prompt set, judge protocol, version identifier, scoring rubric
& Prompt set withheld; rater instructions absent; threshold rationale missing \\[2pt]
Alignment reduction
& ``RLHF reduces toxic outputs 80\% vs.\ base model''
& Training config, reward model, evaluation pipeline, base checkpoint
& Reward model and pipeline almost never released \\[2pt]
Red-team summary
& ``External red team found no CBRN uplift above threshold''
& Protocol, uplift threshold and rationale, test-set scope
& Protocol and threshold rationale almost never published \\[2pt]
Controllability
& ``Constitutional AI produces refusal on 99\% of jailbreak variants''
& Jailbreak set, constitution text, evaluation rubric, checkpoint
& Jailbreak set embargoed; rubric absent \\[2pt]
Risk-card summary
& ``Model poses low risk to critical infrastructure per our evaluation''
& Evaluation scope, methodology, rater qualifications, scoring criteria
& Methodology vague; no independent access; internal-only testers \\
\bottomrule
\end{tabularx}
\end{table}

% ============================================================
\section{The Evidential Inversion: An Evaluation-Methodology Failure}
\label{sec:problem}

\subsection{What the current gap looks like}

NeurIPS has already treated reproducibility as a field-level
responsibility \citep{pineau2021}; the proposal here asks NeurIPS to
apply that same logic to a class of claims that has outpaced existing
norms. Frontier AI safety claims sit at the intersection of two
opposing pressures. They justify release decisions, regulatory
compliance, and public assurances; yet the artefacts needed to
evaluate them (prompts, elicitation procedures, training-data
disclosures, checkpoint access, dangerous-capability test harnesses)
may themselves be sensitive. The recurring result is partial
transparency without scientific reproducibility: provider documents
typically report scores, summaries, or selective examples, but not
enough material for an independent researcher to rerun the evaluation
or stress-test its validity \citep{wan2025fmti,reuel2025frontier}.

Current practice is structurally mixed. Anthropic's April 2026
Responsible Scaling Policy \citep{anthropic2026rsp} commits to public
risk reports with minimised redactions and external reviewers
receiving private versions. OpenAI's Preparedness Framework and recent
system cards \citep{openai2025preparedness} describe tracked
severe-risk categories and external evaluators including CAISI and the
UK AISI. METR has published common-elements analyses across twelve
frontier safety policies \citep{metr2025common}, and
\citet{anthropic2026rsp} report using a fixed index of approximately
100{,}000 human-verified web documents in one evaluation specifically
to enable reproducible evaluation. The core difficulty is not that
reproducibility is impossible; it is that reproducibility is
currently optional, fragmented, and unstandardised.

The April 2026 Claude Mythos Preview release is one concrete
illustration of this gap: a public technical report
\citep{anthropic2026mythos}, an updated RSP risk assessment
\citep{anthropic2026rsp}, and independent evaluations by AISI
\citep{ukaisi2026mythos} and CETaS \citep{hicks2026cetas} all exist,
yet key risk-report sections remain redacted, capability evaluations
rely substantially on internal methodology, and third-party
reproducibility of the safety-relevant claims is still structurally
unavailable. The diagnostic value of the case is precisely that it
occurs under the \emph{most} cooperative disclosure regime
currently observable.

\subsection{Why this is an evaluation-methodology failure}

The evidential inversion is not a transparency preference but a
\emph{measurement-validity} failure with concrete technical evidence.
\citet{chouldechova2025asr}, in the NeurIPS 2025 Position Paper Track,
show that ASR comparisons across systems are often founded on
apples-to-oranges comparisons or low-validity measurements;
``measurement choices, not attack quality, often drive the reported
differences.'' \citet{panfilov2025strategic} demonstrate that frontier
models can generate ``strategic dishonesty'' responses that fool every
output-based jailbreak monitor tested, ``rendering benchmark scores
unreliable.'' \citet{black-box-2024} argue that black-box access is
insufficient for rigorous AI audits. The 2026 IAISR \citep{iaisr2026}
concludes that ``reliable pre-deployment safety testing has become
harder to conduct'' and that models now ``distinguish between
evaluation and deployment contexts and alter their behaviour
accordingly.'' Each of these is a measurement-validity finding, not a
governance complaint. Reproducibility is the standard mechanism by
which such distortions are identified and corrected; the present
proposal applies that mechanism to frontier safety claims.

% Table 2: Evidential inversion -- descriptive vs normative claims (polished caption)
\begin{table}[ht]
\caption{The evidential inversion. As claim consequence rises, reproducibility falls; the disclosure gap is concentrated where claims are normative (\emph{the model is safe enough}) rather than descriptive (\emph{the model achieves $X$ on a benchmark}). Transparency assessments draw on \citet{wan2025fmti} and \citet{fli2025}; the descriptive/normative distinction follows the measurement-theoretic framing in \citet{chouldechova2025asr}.}
\label{tab:inversion}
\centering\small
\begin{tabularx}{\linewidth}{@{}p{2.5cm} p{1.9cm} X p{1.4cm} p{1.4cm}@{}}
\toprule
\textbf{Claim type} & \textbf{Consequence} & \textbf{Current reproducibility} & \textbf{Claim form} & \textbf{Status} \\
\midrule
Standard ML (e.g.\ ImageNet)
& Low--moderate
& High: code, weights, fixed test set
& Descriptive
& Sufficient \\[2pt]
Capability benchmark (e.g.\ MMLU)
& Moderate
& Moderate: prompt set often released
& Descriptive
& Partial gap \\[2pt]
Safety evaluation (e.g.\ HarmBench pass rate)
& High
& Low: summary statistic only
& Normative
& \textbf{Inversion} \\[2pt]
Red-team summary (e.g.\ CBRN uplift)
& Very high
& Minimal: prose; no protocol
& Normative
& \textbf{Inversion} \\[2pt]
Systemic risk (e.g.\ model-card conclusion)
& Governance-defining
& Near zero: no traceable methodology
& Normative
& \textbf{Inversion} \\
\bottomrule
\end{tabularx}
\end{table}

\subsection{Governance consequences}

The consequences are not abstract. The EU AI Act's Article 55 requires
systemic-risk providers to evaluate against ``standardised protocols
\ldots reflecting the state of the art''
\citep[Art.\ 55]{euaiact2024}; the EU AI Office is treating
peer-reviewed research as input to standardisation
\citep{euaioffice2025}. The Delhi Declaration commits signatories to
publish aggregated insights into AI usage \citep{delhi2026};
IASEAI'26 \citep{iaseai2026} convened researchers around
interpretability, oversight, and external evaluation. Across these
venues, the recurring finding is that current frontier safety evidence
is insufficient to support the conclusions drawn from it. NeurIPS can
address the scientific half of that gap by requiring reproducible
evidence before claims acquire the legitimacy of peer-reviewed
publication.

% ============================================================
\section{Why NeurIPS is the Right Site}
\label{sec:why}

\paragraph{NeurIPS already sets reproducibility norms}
The ML Reproducibility Checklist (2019) raised the baseline for code
and data transparency \citep{pineau2021}; the mandatory ethics
statement (2020) established that social consequences are within scope
of review \citep{ashurst2022}; the Datasets and Benchmarks track
(2021) created dedicated infrastructure for evaluation artefacts; the
Position Paper Track (2025) established that normative arguments are
within scope \citep{neurips2025posttrack}. Each was subsequently
adopted by ICML, ICLR, and ACL. Reproducibility standards for frontier
safety claims are the natural next extension.

\paragraph{NeurIPS is where frontier safety claims acquire scientific legitimacy}
System cards for major frontier models cite NeurIPS-published research;
alignment benchmark papers
\citep{lin2022truthfulqa,mazeika2024harmbench,zhao2024wildguard} are
NeurIPS-family; and the dominant alignment methods (RLHF
\citep{ouyang2022}, Constitutional AI \citep{bai2022constitutional},
and DPO \citep{rafailov2023}) were published at or around NeurIPS.

\paragraph{Direct continuity with 2025 NeurIPS position-paper precedent}
The proposal extends five accepted 2025 NeurIPS position papers.
\citet{bommasani2025consensus} argued NeurIPS should lead consensus on
AI policy; we argue NeurIPS should require a specific reproducibility
standard as one input to that consensus.
\citet{schaeffer2025refutations} proposed a Refutations \& Critiques
track for post-hoc correction; reproducibility standards reduce the
volume of claims requiring post-hoc critique by making them
scrutinisable ex-ante. \citet{chouldechova2025asr} showed ASR
comparisons require valid measurement; our framework operationalises
that finding for review-time. \citet{dolin2025postdep} argued for
statistically valid post-deployment monitoring in clinical AI; we
argue the analogous standard for pre-deployment frontier safety
claims. \citet{olteanu2025rigor} proposed a six-pillar view of rigor;
the present proposal directly operationalises \emph{reporting rigor}
for the highest-stakes claim category. Appendix~\ref{app:precedent}
makes this comparison explicit.

\paragraph{The ACM artefact-review precedent}
ACM's badging framework \citep{acm2020artifact} does not assume every
paper can publicly release every artefact; it distinguishes
\emph{available, reviewed}, and \emph{validated} statuses. Proprietary
artefacts may be shown to reviewers without public release. This
graduated signalling is the right model for frontier safety claims:
not a binary release/trust choice, but a structured statement of what
is available, to whom, under what constraints, and with what
implications for claim strength.

% ============================================================
\section{The Proposal: Three-Tier Disclosure and Claim Inventory}
\label{sec:proposal}

\subsection{The claim inventory}

NeurIPS should require a \emph{claim inventory} for any paper making
frontier AI safety claims. Authors list each such claim, state its
scope, identify the artefacts needed to evaluate it, declare the
requested disclosure tier, and specify any uncertainty introduced by
access constraints, checkpoint drift, deployment changes, elicitation
choices, or distribution shift between evaluation and deployment
settings. The inventory is modelled on the 2020 ethics statement
\citep{ashurst2022} and aligned with the STREAM standard for
transparently reporting evaluations \citep{mccaslin2025stream}.

\subsection{The three-tier framework}

% Table 3: Three-tier disclosure framework
\begin{table}[ht]
\caption{Three-tier disclosure framework for frontier AI safety claims. T1 is the default. T3 is explicitly claim-limiting: a paper whose core empirical contribution depends on a T3 claim is not suitable as an empirical frontier-safety paper at NeurIPS.}
\label{tab:tiers}
\centering\small
\begin{tabularx}{\linewidth}{@{}p{1.6cm} p{2.6cm} X X@{}}
\toprule
\textbf{Tier} & \textbf{When it applies} & \textbf{Minimum evidential requirement} & \textbf{Claim strength supported} \\
\midrule
\textbf{T1}\par\textcolor{tier1}{Public}
& Public release does not materially raise misuse, security, privacy, or legal risk
& Public release of evaluation prompts, code, rubrics, datasets or proxies, version/checkpoint identifiers, and instructions sufficient to rerun the claim
& Full ordinary empirical support \\[3pt]
\textbf{T2}\par\textcolor{tier2}{Controlled}
& Public release would create material risk; confidential review is feasible
& Private access for a NeurIPS confidential panel or approved independent auditors. Public: summary of methods and results, redaction justifications, reviewer scope statement
& Strong but qualified empirical support \\[3pt]
\textbf{T3}\par\textcolor{tier3}{Restricted}
& Even confidential review is not feasible (e.g.\ live CBRN uplift harnesses)
& Public statement of: claim, reasons for restriction, why neither T1 nor T2 is possible. Claim must be right-scaled to what the restricted evidence can actually support
& Limited contextual or descriptive use only; not sufficient for threshold-crossing or release-justifying claims \\
\bottomrule
\end{tabularx}
\end{table}

The critical design principle for Tier~3 is \emph{right-scaling the
claim}: if authors cannot share artefacts even under confidentiality,
the associated claim should not disappear from the paper but it loses
the right to function as a strong scientific warrant. A T3 claim may
appear as context, operational experience, or a restricted-disclosure
note, but not as the basis for ``below threshold,'' ``mitigations
adequate,'' or ``release supported by evaluation evidence.''

\paragraph{Tier classification is reviewable}
A first-order failure mode of any tiered framework is gaming: authors
classifying a claim at a lower tier than its sensitivity warrants, or
at a higher tier than necessary to avoid the burden of public release.
Tier selection is therefore reviewable in two places. At submission,
ordinary reviewers verify that the requested tier is consistent with
the claim form (Table~\ref{tab:checklist}, S2): a normative claim
about a deployment-relevant threshold cannot be classified T3 without
an explicit right-scaling statement. At Phase~2, the confidential
panel verifies that artefacts in T2 are genuinely sensitive enough to
warrant controlled rather than public disclosure, and may downgrade.
The full reviewability protocol is in Appendix~\ref{app:governance}.

\paragraph{T2 review need not always require a full panel}
A single domain-matched reviewer may sign a public adequacy statement
in place of the full panel for T2 claims whose artefacts contain
\emph{no} CBRN-uplift, cyber-offence, deception, or
systemic-risk-threshold content (e.g., large evaluation prompt sets
without CBRN material, proprietary but non-hazardous evaluation
pipelines); the full 5--7 person panel is required whenever any of
these categories is present (Appendix~\ref{app:governance}, Section~D.5).
This rule prevents mission creep: claims with the largest expected
governance impact (systemic-risk threshold-crossing,
self-replication, autonomous cyber-offence) cannot be routed through
the lighter pathway.\footnote{The general pattern (a small confidential panel reviewing artefacts that cannot be released publicly) is well-established outside ML: medical research ethics review boards review patient-identifiable data without public release; classified peer review handles security-sensitive scientific work for defense-adjacent agencies; coordinated vulnerability disclosure mechanisms operate via CERT teams that examine exploit details before any public disclosure. The novelty here is the venue (a scientific conference) and the trigger (peer-reviewed publication legitimacy), not the mechanism of confidential expert review itself.}

\subsection{Mandatory scope statements}

The claim inventory must include a scope statement specifying:
(a)~\emph{what system was evaluated} (training checkpoint,
representative launch checkpoint, refusal-relaxed checkpoint,
deployment proxy, or production configuration); (b)~\emph{what
evaluation environment was used} (offline, live deployment, or
protected parallel environment); (c)~\emph{what pipeline components
were present} (elicitation, scoring, monitoring, post-processing,
human escalation); and (d)~\emph{what uncertainty applies}
(distribution shift between test and deployment, judge-model error
rates, evaluation-awareness effects;
\citealp{iaisr2026,panfilov2025strategic}). Scope-statement failure
makes the corresponding claim \emph{scope-limited} regardless of the
disclosure tier.

% Figure 1: Three-tier disclosure flow (TikZ)
% Layout: q2 sits directly above t2 so the Yes arrow drops straight down
% into Tier 2; t1 sits far-left below q1 (Yes from q1.west); t3 sits
% far-right of q2 (No from q2.east). This mirrors the q1 geometry: Yes goes
% one way (here, straight down), No goes the other (right then down).
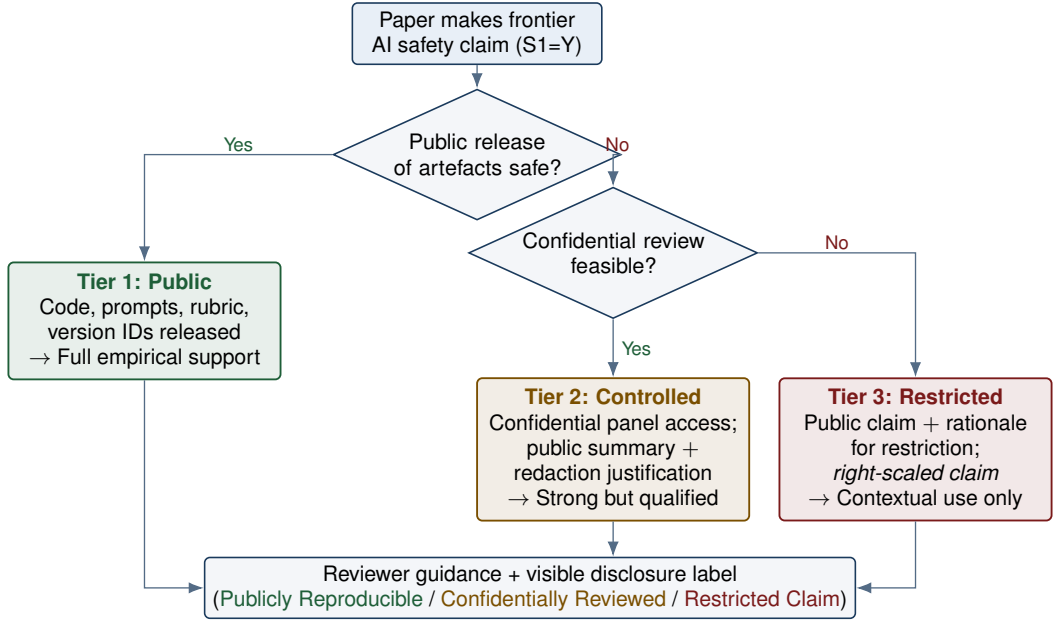
\begin{figure}[ht]
\centering
\begin{tikzpicture}[
  font=\sffamily\footnotesize,
  process/.style={
    rectangle, rounded corners=2pt, draw=neuripsnavy, line width=0.6pt,
    fill=lightnavy, minimum height=8mm, minimum width=33mm, align=center,
    inner sep=3pt
  },
  decision/.style={
    diamond, draw=neuripsnavy, line width=0.6pt, fill=boxgray,
    aspect=2.2, minimum height=8mm, align=center, inner sep=1pt
  },
  tier1box/.style={
    rectangle, rounded corners=2pt, draw=tier1, line width=0.7pt,
    fill=tier1!10, minimum height=14mm, minimum width=36mm, align=center,
    inner sep=4pt
  },
  tier2box/.style={
    rectangle, rounded corners=2pt, draw=tier2, line width=0.7pt,
    fill=tier2!10, minimum height=14mm, minimum width=36mm, align=center,
    inner sep=4pt
  },
  tier3box/.style={
    rectangle, rounded corners=2pt, draw=tier3, line width=0.7pt,
    fill=tier3!10, minimum height=14mm, minimum width=36mm, align=center,
    inner sep=4pt
  },
  arrow/.style={-{Latex[length=2mm]}, line width=0.5pt, draw=neuripsnavy!75},
  yeslab/.style={font=\sffamily\scriptsize, text=tier1, inner sep=1pt},
  nolab/.style={font=\sffamily\scriptsize, text=tier3, inner sep=1pt},
]

\coordinate (centre) at (0.5, 0);

\node[process] (paper) at ($(centre)+(0,0)$) {Paper makes frontier\\AI safety claim (S1=Y)};

\node[decision, text width=24mm] (q1) at ($(centre)+(0,-1.6)$) {Public release\\of artefacts safe?};

\node[decision, text width=24mm] (q2) at ($(centre)+(1.8,-2.9)$) {Confidential review\\feasible?};

\node[tier1box] (t1) at ($(centre)+(-4.4,-3.8)$)
  {\textbf{\textcolor{tier1}{Tier 1: Public}}\\Code, prompts, rubric,\\version IDs released\\$\to$ Full empirical support};

% Tier 2 directly below q2 -- gives a straight vertical Yes arrow
\node[tier2box] (t2) at ($(centre)+(1.8,-5.5)$)
  {\textbf{\textcolor{tier2}{Tier 2: Controlled}}\\Confidential panel access;\\public summary $+$\\redaction justification\\$\to$ Strong but qualified};

\node[tier3box] (t3) at ($(centre)+(5.8,-5.5)$)
  {\textbf{\textcolor{tier3}{Tier 3: Restricted}}\\Public claim $+$ rationale\\for restriction;\\\emph{right-scaled claim}\\$\to$ Contextual use only};

\path (t1.south) -- (t3.south) coordinate[midway] (midbottom);
\node[process, below=14mm of midbottom, minimum width=86mm, fill=boxgray]
  (label)
  {Reviewer guidance + visible disclosure label\\(\textcolor{tier1}{Publicly Reproducible}~/~%
  \textcolor{tier2}{Confidentially Reviewed}~/~%
  \textcolor{tier3}{Restricted Claim})};

\draw[arrow] (paper) -- (q1);
\draw[arrow] (q1.west) -| node[yeslab, pos=0.25, above] {Yes} (t1.north);
\draw[arrow] (q1.east) -| node[nolab, pos=0.25, above] {No} (q2.north);
% Straight-down Yes arrow (q2 is directly above t2)
\draw[arrow] (q2.south) -- node[yeslab, pos=0.5, right=2pt] {Yes} (t2.north);
\draw[arrow] (q2.east) -| node[nolab, pos=0.25, above] {No} (t3.north);

\draw[arrow] (t1.south) |- (label.west);
\draw[arrow] (t2.south) -- (t2.south |- label.north);
\draw[arrow] (t3.south) |- (label.east);

\end{tikzpicture}
\caption{The three-tier disclosure decision flow. A paper that makes a
frontier AI safety claim (S1$=$Yes on the proposed checklist) is routed by
two questions: (1)~can the artefacts be released publicly without raising
misuse risk? and (2)~if not, can a qualified confidential panel review them?
T1 is the default; T2 covers genuine cases of justified secrecy; T3 exists
but is \emph{claim-limiting}. Each accepted paper carries a visible
disclosure label modelled on ACM's artefact badges \citep{acm2020artifact}.}
\label{fig:flow}
\end{figure}

\subsection{Potential pitfalls of the framework itself}
\label{sec:pitfalls}

A serious proposal must anticipate that it might itself fail. Beyond
the tier-gaming concern addressed above (Section~5.2), three pitfalls
remain.

\paragraph{Confidential-panel capture}
The pool of qualified frontier-safety reviewers is small and partially
overlapping with the labs being audited. Mitigation: panel membership
is restricted to \emph{unaffiliated academic researchers and personnel
of national AI Safety Institutes}, with three-year fixed terms,
mandatory recusal on prior collaboration within five years, and
auditable access logs (Appendix~\ref{app:governance}).

\paragraph{Threshold-anchor obsolescence}
The compute-based frontier anchor (Section~\ref{sec:def}) may become stale
as training efficiency improves. Mitigation: the anchor admits a
capability-based alternative (emergent dangerous capabilities), and
the NeurIPS programme committee is asked to revisit thresholds
annually, following the precedent of the reproducibility checklist's
iterative revision.

\paragraph{T2 review at conference scale}
Even a small Phase~2 pilot requires secure infrastructure and legal
agreements that NeurIPS does not currently possess. Mitigation: the
Phase~2 pilot is bounded to at most five papers per cycle, and T2
review is hosted via a \emph{federated colloquium of secure-review
entities}: national AI Safety Institutes (UK AISI, US AISI, Japan
AISI, and analogous bodies emerging from the EU AI Office and the
India AI Mission), independent evaluators (METR, Centre for AI
Safety), and trusted academic secure-review labs, with the host
chosen by claim domain rather than geography. Properties (parallel
capacity, political robustness, domain matching, institutional
humility) and the allocation protocol are elaborated in
Appendix~\ref{app:federated}.

% ============================================================
\section{Implementation}
\label{sec:impl}

\paragraph{Phase 1: Claim inventory and checklist (Year 1)}
NeurIPS adds the five-question frontier safety claim section (Table~\ref{tab:checklist}) to the author checklist; the gating question (S1) takes minutes for the vast majority of papers and full claim inventories are required only for papers where S1=Yes.

Authors of papers with S1$=$Yes complete the claim inventory at
submission. Ordinary reviewers assess the inventory under guidelines
extending the 2024 NeurIPS Ethics Review process \citep{ashurst2022};
no new panel is required. \textbf{This is the minimum viable
adoption}: claim inventory, scope statement, and disclosure tier on
the existing checklist, with no confidential infrastructure or new
legal agreements. Reviewers are not expected to validate
specialised CBRN, cyber, or deception evaluations unaided;
Phase~1 checks that the inventory and scope statement are present
and coherent, escalating to domain reviewers (or the confidential
panel of Appendix~\ref{app:federated}) where claim validity turns on
specialised expertise.

\paragraph{Phase 2: Confidential frontier-claims panel (Year 2 pilot)}
NeurIPS pilots a small confidential panel, bounded to \textbf{at most
five papers} per cycle, modelled on third-party compliance review
\citep{thirdparty2025}. The panel comprises 5--7 unaffiliated academic
researchers and AISI personnel under fixed three-year terms with
mandatory five-year COI recusal. T2 artefacts are received under
NDA-equivalent confidentiality with auditable access logs. The panel
produces a public adequacy statement; ordinary reviewers continue to
assess the scientific contribution. T2 review is hosted via the
federated colloquium of Appendix~\ref{app:federated}; host
assignment is a function of claim domain
(Appendix~\ref{app:governance}).

\paragraph{Phase 3: Disclosure labels, sanctions, and cross-conference adoption (Year 3)}
Accepted papers receive a visible disclosure label
(\emph{Publicly Reproducible}, \emph{Confidentially Reviewed}, or
\emph{Restricted Claim}) mirroring ACM's artefact badges
\citep{acm2020artifact}. \textbf{Graduated sanctions} apply for
non-compliance: tier-justification inadequate $\to$ revision request;
tier downgrade refused by authors $\to$ claim-strength reduction in
metareview; submission found to have claimed T2 without good-faith
justification $\to$ desk rejection and referral to senior PC. NeurIPS
publishes aggregate statistics on tier distribution. ICML and ICLR
are invited to co-adopt, consistent with the Reproducibility
Checklist's diffusion path \citep{pineau2021}.

% Table 4: Proposed NeurIPS author checklist additions
\begin{table}[t!]
\caption{Five proposed additions to the NeurIPS author checklist. S1 is the gate: only papers where S1$=$Yes face S2--S5. Estimated author completion time: S1 screening takes minutes; full claim inventories for S1$=$Yes papers may require several hours, proportional to claim complexity.}
\label{tab:checklist}
\centering\small
\begin{tabularx}{\linewidth}{@{}p{0.6cm} p{4.6cm} X@{}}
\toprule
\textbf{Q} & \textbf{Question} & \textbf{Reviewer guidance} \\
\midrule
S1 & Does this paper make a frontier AI safety claim? (Y / N / NA)
   & If Y, S2--S5 apply. If authors answer N but the paper contains safety-relevant claims about a frontier model, flag for AC attention. \\[3pt]
S2 & For each safety claim: which disclosure tier applies, and what is the justification? (T1 / T2 / T3 with written rationale)
   & Tier selection is reviewable. T2 or T3 requires written justification. Unsupported T3 without claim-limiting language should be flagged as non-compliance. \\[3pt]
S3 & Have evaluation prompts, judge specifications, rubric, and version/checkpoint identifiers been provided under the requested tier? (Y / N + explanation)
   & N without a T2/T3 justification should be treated as unsupported evidence and noted in the review. \\[3pt]
S4 & For each claim: what is the scope? (checkpoint type / evaluation environment / pipeline components; see author guidelines)
   & Missing or vague scope statements should trigger a request for revision. Claims that do not specify offline vs.\ production setting are scope-limited. \\[3pt]
S5 & Are confidence intervals or uncertainty quantifications reported for all numerical safety claims? (Y / N + explanation)
   & Point estimates without uncertainty quantification should be flagged; recommend revision. \\
\bottomrule
\end{tabularx}
\end{table}

% ============================================================
\section{Alternative Views}
\label{sec:alt}

\paragraph{``Safety-sensitive information cannot be released.''}
The most important safety evaluations (CBRN uplift
\citep{vaccaro2026uplift}, cyber-offence capability,
critical-infrastructure risk) cannot be released because doing so
creates safety risks; a reproducibility mandate is incompatible with
responsible safety evaluation. \textbf{Response:} The three-tier framework
addresses this directly: T2 and T3 do not require public release. The
relevant precedent is confidential expert review (medical research
ethics boards, classified peer review, coordinated vulnerability
disclosure), not regulatory compliance: sensitive details
scrutinised by a small, trusted group before a public claim
receives peer-reviewed legitimacy. NeurIPS does not become a regulator; it does not certify
that a model is safe, only the evidentiary status of claims made in
papers it publishes. The prestige of a NeurIPS-vetted safety claim
is the incentive that aligns labs with the standard rather than
against it. The question is not whether sensitive artefacts can be
published but whether they can be scrutinised by a qualified
independent party \citep{thirdparty2025,reuel2025frontier}. We
argue these artefacts must be scrutinised in this way, or the claim
must be narrowed.

\paragraph{``Frontier labs do not need NeurIPS legitimacy for governance purposes.''}
Frontier AI labs can publish through corporate channels, government
partnerships, or non-peer-reviewed reports; NeurIPS legitimacy is not
a precondition for governance impact. \textbf{Response:} In practice,
labs submit their most significant technical work to NeurIPS because
peer review confers legitimacy that corporate or governmental
publication does not, and regulators themselves distinguish
peer-reviewed findings from non-peer-reviewed assertions
\citep{bommasani2025consensus,euaioffice2025}. The proposal covers
exactly those claims for which authors seek peer-reviewed
legitimacy; governance-relevant claims should meet a
governance-relevant evidentiary standard.

\paragraph{``This would deter industry participation.''}
Stricter standards would push frontier labs toward non-peer-reviewed
channels. \textbf{Response:} Current frameworks already assume mixes
of public reporting, redaction, and trusted-actor disclosure
\citep{anthropic2026rsp,seoul2024,openai2025preparedness}. The
proposal asks labs to map that practice onto a conference-standard
format, or accept narrower claims. If a lab cannot satisfy even T2
disclosure for a safety claim, that is information about the claim,
not a reason to exempt it from scrutiny.

\paragraph{``Reproducibility is undermined by deployment-context shift.''}
Reproducibility presupposes a fixed evaluation context, but frontier
safety operates under distribution shift, adversarial prompting, and
deployment mismatch. \textbf{Response:} Addressed by the mandatory scope
statement (Section~5.3): claims about deployment from offline evaluation
are explicitly scope-limited. Reproducibility does not eliminate
deployment-context uncertainty; it makes that uncertainty
\emph{traceable}, the precondition for reducing it
\citep{iaisr2026}.

\paragraph{``Confidential review is vulnerable to capture or leakage.''}
The confidential panel is vulnerable to COI capture and to leakage of
sensitive artefacts. \textbf{Response:} The risk is real. The response is
governance not waiver: panel restricted to unaffiliated academics and
AISI personnel, fixed terms, five-year COI recusal, auditable access
logs, public tier rationales \citep{thirdparty2025,mokander2023}.
Anthropic's external-review design already operates under this logic
\citep{anthropic2026rsp}; detailed protocol in
Appendix~\ref{app:governance}.

% ============================================================
\section{Limitations and Conclusion}

The proposal
above is a phased scientific-review standard, not a ready-made
regulatory infrastructure: Tier~2 controlled review is
capacity-constrained and should begin only as a bounded pilot, after
the Year-1 baseline (claim inventory, scope statement, and
claim-strength labelling) has been tested in practice. The most consequential claims in AI safety are often the least
reproducible. This is not a quirk of the field; it is a structural
consequence of applying disclosure norms calibrated for conventional
ML claims to a class of claims that is consequentially different in
kind: claims that justify deployment and shape governance of systems
whose behaviour the public cannot directly observe.  NeurIPS does
not need to become a regulator to fix this; it needs only to apply,
to its highest-stakes claims, the standard it already applies to its
lowest. The time to make that standard explicit is now, while the
framework can still shape practice, rather than chase it.

\clearpage
\bibliographystyle{plainnat}
\bibliography{refs}

% ============================================================
\clearpage
\appendix
\section{Notes and Definitions}

\textbf{Note 1.} \emph{Frontier vs.\ AGI.} The EU AI Act, UK AI Safety
Framework, and NIST AI RMF organise obligations around
``general-purpose AI models with systemic risk'' and ``frontier
models'', both operationally defined categories, rather than
``AGI.''

\textbf{Note 2.} The ML Reproducibility Checklist was introduced at
NeurIPS 2019 following Joelle Pineau's 2018 keynote and was
subsequently adopted by ICML, ICLR, and ACL.

\textbf{Note 3.} \emph{Right-scaling the claim} means reducing the
strength of the conclusion to what the accessible evidence can
support.

\textbf{Note 4.} \emph{Qualified auditor} for T2: an entity with
domain expertise, independence from the submitting organisation, and
an established process for receiving and protecting sensitive
technical information. Current candidates include national AI Safety
Institutes (UK, US, Japan, and analogous bodies emerging from the EU
AI Office and the India AI Mission), METR, and the Centre for AI
Safety; the federated colloquium framing of
Appendix~\ref{app:federated} treats this set as composable rather
than singular.

\clearpage
\section{Illustrative Tier-2 Submission}
\label{app:illustrative}
% Appendix A: Illustrative Tier-2 submission
% Modelled on Schaeffer et al. (2025)'s illustrative example approach

We provide a redacted, illustrative example of what a T2 submission for
a CBRN red-teaming claim would contain in practice. The example is
synthetic; it is intended to make the framework concrete rather than to
endorse any specific result.

\paragraph{Hypothetical paper title}
``Constitutional Refusal Training Reduces CBRN Uplift on Frontier-7B
Below Critical Capability Threshold.''

\paragraph{Claim inventory (excerpt)}
\begin{itemize}
\item \textbf{C1 (T2)}: ``Frontier-7B with constitutional training does
not exceed CCL-3 (medium uplift) on internal CBRN red-team set,
$N=2{,}400$ prompts, panel-of-judges adjudication.''
\item \textbf{C2 (T1)}: ``Constitutional training reduces refusal-rate
on the public HarmBench-CBRN-public subset from 71\% to 94\%,
$\Delta = 23$ percentage points (95\% CI [19.4, 26.7]).''
\item \textbf{C3 (T3 / right-scaled)}: ``Operational deployment
monitoring over 90 days suggests no incident escalation; \emph{this is
context, not a release-justifying warrant}.''
\end{itemize}

\paragraph{Scope statement (C1)}
System: launch-candidate checkpoint $v3.2.1$, refusal-relaxed for
elicitation, full safety stack disabled. Environment: offline
evaluation. Pipeline components: zero-shot elicitation only (no agent
scaffolding). Uncertainty: judge-model error rate not estimated;
distribution shift between offline and deployment unmeasured;
evaluation-awareness effects \citep{iaisr2026,panfilov2025strategic} not
controlled for.

\paragraph{T2 disclosure package (provided to confidential panel)}
Full prompt set with provenance; judge-model prompts and rubrics;
threshold definition and rationale relative to CCL-3; per-prompt
scoring; calibration-set kappa scores; checkpoint hash and access
instructions; raw activation logs for sampled prompts.

\paragraph{Public summary}
``On a fixed 2{,}400-prompt CBRN evaluation reviewed under T2, the
panel found the threshold-non-crossing claim adequately supported under
the stated scope, with three minor adequacy gaps documented in the
panel's adequacy statement (Appendix~B of the published version). The
underlying prompt set is restricted under T2 confidentiality;
qualified-researcher access available via \texttt{[procedure]}.''

\paragraph{Panel adequacy statement (excerpt)}
``Claim C1 is adequately supported within the stated scope. The panel
notes three caveats: (i)~the judge-model used has $\sim$8\% disagreement
with majority human adjudication on a 200-prompt calibration sample,
which is acknowledged but not propagated into the threshold
calculation; (ii)~the elicitation protocol does not include agent
scaffolding, which limits the claim's coverage to non-agentic
deployment; (iii)~no evaluation-awareness ablation was performed.
Authors should add a corresponding scope-limitation statement.''

\paragraph{What this example shows}
The framework allows the consequential claim (C1) to be evaluated
without public release of the prompt set; the auxiliary descriptive
claim (C2) is fully reproducible under T1; and the operational
observation (C3) is preserved as context but cannot bear the weight of
a release decision. This is the work-product the framework is designed
to produce.

\clearpage
\section{Comparison with 2025 NeurIPS Position-Paper Precedents}
\label{app:precedent}
% Appendix C: Comparison with 2025 NeurIPS position-paper precedents (polished -- Status column removed)

\begin{table}[ht]
\caption{Comparison with five 2025 NeurIPS Position Paper Track exemplars. All five address some aspect of evidence quality, claim validity, or institutional norm-setting; the present proposal extends each in a complementary direction.}
\label{tab:precedent}
\centering\small
\begin{tabularx}{\linewidth}{@{}p{3.4cm} X@{}}
\toprule
\textbf{Precedent paper} & \textbf{Relation to present proposal} \\
\midrule
\citet{bommasani2025consensus} ``NeurIPS should lead scientific consensus on AI policy''
& Argues NeurIPS should orchestrate consensus formation. Present paper proposes a specific reproducibility standard as one of the technical inputs such consensus would require. \\[3pt]
\citet{schaeffer2025refutations} ``Refutations and Critiques Track''
& Proposes a track for post-hoc correction of flawed papers. Present paper acts upstream: stronger reproducibility standards for frontier safety claims reduce the volume of claims requiring later refutation by making them scrutinisable ex-ante. \\[3pt]
\citet{chouldechova2025asr} ``Comparison requires valid measurement'' (red-teaming ASRs)
& Shows that ASR comparisons across systems are often founded on apples-to-oranges measurements. Present paper operationalises this finding by requiring scope statements, judge-model error reporting, and threshold rationales as conditions of T1/T2 disclosure. \\[3pt]
\citet{dolin2025postdep} ``Statistically Valid Post-Deployment Monitoring'' for clinical AI
& Argues post-deployment monitoring should be standard in clinical AI, with statistically valid testing frameworks. Present paper makes the analogous argument for pre-deployment frontier safety claims, with reproducibility tiers as the analog of the post-deployment statistical guarantees. \\[3pt]
\citet{olteanu2025rigor} ``Rigor in AI'' (six-pillar framework)
& Identifies methodological, epistemic, normative, conceptual, reporting, and interpretative rigor as distinct dimensions. Present paper directly operationalises \emph{reporting rigor} for the highest-stakes claim category at NeurIPS. \\
\bottomrule
\end{tabularx}
\end{table}

\paragraph{Pattern across the precedents}
All five address a common underlying concern: the evidence on which AI
claims acquire scientific legitimacy is structurally weaker than the
strength of the claims being made. Bommasani's ``consensus gap''
(presented as Oral at NeurIPS 2025), Schaeffer et al.'s
``self-correction gap'' (also Oral), Chouldechova et al.'s
``measurement-validity gap,'' Dolin et al.'s ``post-deployment
surveillance gap,'' and Olteanu et al.'s ``narrow-rigor gap'' all
describe variants of the same pathology: claims outpace the evidence
that licenses them. The present proposal addresses the specific case
where this pathology is most acute (frontier AI safety claims)
with a specific mechanism (tiered disclosure with right-scaling).

\paragraph{Why this matters for track-fit}
The NeurIPS 2026 Position Paper Track explicitly broadens scope toward
``technically grounded'' positions \citep{neurips2026track}. The five
precedents above are all technically grounded: they ground their
normative claims in measurement theory, statistical inference,
sociotechnical analysis, or institutional design. The present paper
follows the same template: the central diagnostic claim (the evidential
inversion) is grounded in measurement-validity findings
\citep{chouldechova2025asr,panfilov2025strategic,black-box-2024}, and
the proposal is a concrete review-time mechanism rather than a
governance call.

\clearpage
\section{Tier-2 Reviewer Governance Protocol}
\label{app:governance}
% Appendix D: Tier-2 reviewer governance protocol (polished)

This appendix specifies the governance design for the Phase~2 Tier-2
confidential panel. The design draws on
\citet{thirdparty2025},
\citet{mokander2023}, and the NeurIPS Ethics Review process
\citep{ashurst2022}.

\paragraph{D.1 Panel composition}
The Tier-2 Confidential Panel comprises 5--7 members serving fixed
three-year terms. Membership is restricted to:
\begin{itemize}
\item Tenured or tenure-track academic researchers in ML, AI safety,
or measurement theory, with no current funding from a frontier AI lab
within the prior 24 months;
\item Personnel of national AI Safety Institutes (UK AISI, US AISI,
Japan AISI, or successor entities);
\item Researchers from established third-party evaluation
organisations (e.g., METR, Centre for AI Safety) provided that the
organisation does not receive a majority of its funding from any
single frontier developer.
\end{itemize}
Industry researchers may serve as \emph{non-voting consultants} on
specific submissions when domain expertise is unavailable from voting
members; consultant input is documented in the panel's adequacy
statement.

\paragraph{D.2 Conflicts of interest}
Mandatory recusal applies on any of: prior co-authorship within five
years; current or recent (within 24 months) funding relationship with
the submitting organisation; equity holdings exceeding standard index
exposure; immediate family employed at the submitting organisation. COI
disclosures are made in writing at the start of each cycle and
re-affirmed at submission assignment.

\paragraph{D.3 Access and handling of T2 artefacts}
Submitting authors transmit T2 artefacts to a secure-review
infrastructure operated by the partner entity (a national AISI or
equivalent). Panel members access artefacts via authenticated channels
with auditable access logs. Local copies are prohibited; review takes
place in a secure environment provided by the partner. NDAs are
co-signed by NeurIPS, the partner, and each panel member, with
explicit termination clauses tied to publication or rejection of the
submitting paper. Hosts operate under their existing secure-handling
protocols for sensitive technical material (analogous to those used
for medical-data review boards, classified-research panels, and
coordinated vulnerability disclosure), so legal and infrastructural
liability rests with institutions already equipped to bear it rather
than with NeurIPS.

\paragraph{D.4 Preserving double-blind review under T2}
T2 review introduces a tension with NeurIPS double-blind policy:
secure transmission of artefacts requires that the host entity know
the submitting authors. The protocol resolves this by partitioning
identity disclosure: the secure-review host is permitted to know
author identity for credentialing and chain-of-custody purposes, but
the confidential panel reviewing the artefacts works from a
de-identified package containing only the artefacts, the claim
inventory, the scope statement, and a host-issued provenance hash.
The host does not transmit author-identifying metadata to the panel,
and the panel's adequacy statement is signed by panel role rather
than by individual reviewer. Ordinary scientific reviewers (assessing
the paper itself) remain blinded to authorship throughout, as in any
standard NeurIPS submission.

\paragraph{D.5 Scope bounds and a path to scale}
The Phase~2 pilot is bounded to \textbf{at most five papers per cycle},
with the panel meeting twice during the review window: once after
initial review for tier-justification adequacy, and once after revision
to verify that any tier downgrades or scope-statement clarifications
have been incorporated.

The five-paper bound is a deliberate floor, not a ceiling. We propose a
graduated scaling path from Year~3 onwards. If observed T2 demand
exceeds capacity in a given cycle, NeurIPS adopts \emph{selective
sampling}: T2 requests are triaged by the senior PC into (i)~ones
warranting full panel review (capped at the cycle's panel capacity),
(ii)~ones reviewed by a single domain-matched panel member with
written sign-off, and (iii)~ones referred to a partner entity in the
federated colloquium (Appendix~\ref{app:federated}) under its own evaluation
pipeline. Full-panel review is reserved for claims with the largest
expected governance impact (e.g., systemic-risk threshold-crossing
claims). The scaling formula is panel capacity $=$ membership
$\times$ average review hours per member per cycle, with the bound
revised annually based on observed cost and review quality. This
avoids the failure mode in which the panel becomes a bottleneck at
higher T2 volumes.

\paragraph{D.6 Output: the panel adequacy statement}
For each T2 submission, the panel produces a public adequacy statement
of 200--500 words, included in the published paper. The statement
records: (i)~the tier requested and the panel's assessment of its
adequacy; (ii)~any caveats applied to specific claims; (iii)~the
panel's recommendation regarding scope-limitation language;
(iv)~recused members; (v)~the partner entity hosting secure review.
The statement does not reveal artefact content but does record the
panel's confidence in the tier classification.

\paragraph{D.7 Sanctions and dispute resolution}
\textbf{Graduated sanctions}, applied by the senior PC after panel
recommendation:
\begin{itemize}
\item \emph{Tier-justification inadequate}: revision request with
specific guidance.
\item \emph{Tier downgrade refused by authors}: the corresponding
claim is reduced in strength in the metareview, and the published
paper carries a Restricted Claim disclosure label.
\item \emph{T2 claimed without good-faith justification}: desk
rejection and referral to senior PC; in egregious cases, referral to
the NeurIPS Code of Conduct process.
\end{itemize}
Author dispute resolution follows the existing NeurIPS appeals
process, extended with one panel-meeting cycle to allow for rebuttal
on tier classification specifically.

\paragraph{D.8 Cost estimate and funding}
A five-paper Phase~2 pilot is estimated at 200--300 panel-hours
(40--60 hours per submission across 5--7 members), plus secure-review
infrastructure absorbed by the partner entity. The assumed funding
model is (i)~partnership with one or more federated hosts
(Appendix~\ref{app:federated}) for infrastructure and panellist honoraria,
with (ii)~the NeurIPS Foundation as fallback for honoraria if no host
in the colloquium can absorb the cost in a given cycle. The total
cost envelope is comparable to the existing NeurIPS Ethics Review
\citep{ashurst2022} and is well within the operational capacity of any
single national AISI, let alone the colloquium as a whole.

\paragraph{D.9 Why governance, not waiver}
The objections in Section~\ref{sec:alt} all reduce to a question about the
feasibility of governance. Existing institutions
(clinical-trial review boards, financial-sector audit, nuclear-energy
peer review) routinely operate at this complexity. The argument for
waiving reproducibility entirely on feasibility grounds is much weaker
than the argument for governing it carefully.

\clearpage
\section{Comparison with Technical Alternatives}
\label{app:alternatives}
% Appendix D: Comparison with technical alternatives

A reasonable reviewer might ask why tiered reproducibility, rather than
some other technical mechanism, is the right tool. We compare with
three alternatives that have received serious treatment in the
literature: standardised benchmarking, third-party auditing, and
red-teaming infrastructure. Each is valuable; none substitutes for
review-time disclosure standards on the specific category of frontier
safety claims.

\begin{table}[ht]
\caption{Comparison of mechanisms for evaluating frontier AI safety claims at NeurIPS-publication time. ``Review-time'' indicates whether the mechanism operates during scientific peer review (the present paper's concern), as opposed to before submission or after publication.}
\label{tab:alternatives}
\centering\small
\begin{tabularx}{\linewidth}{@{}p{2.4cm} X p{1.8cm} p{1.6cm}@{}}
\toprule
\textbf{Mechanism} & \textbf{What it does well; what it does not} & \textbf{Operates at} & \textbf{Substitute?} \\
\midrule
Standardised benchmarking (e.g.\ HarmBench, WildGuard)
& Provides a fixed evaluation distribution that enables cross-system comparison. But: a fixed benchmark cannot cover novel sensitive evaluations (CBRN uplift on new models, jailbreak protocol unique to a deployment); benchmark contamination is widespread; \citet{chouldechova2025asr} show ASR comparisons across benchmarks are often invalid even with fixed prompt sets.
& Pre-submission
& No, complementary \\[3pt]
Third-party auditing
(e.g.\ METR, AISIs, FMTI methodology)
& Produces independent assessment of specific systems by qualified evaluators. \citet{reuel2025frontier} note current third-party audits suffer from inconsistent scope, access, and reporting. Audits are slow ($\sim$weeks to months), cost-intensive, and address specific systems rather than the general validity of claims published at NeurIPS.
& Post-submission / continuous
& No, complementary \\[3pt]
Red-teaming infrastructure
(internal + external red teams; secure enclaves)
& Produces evidence about specific vulnerabilities. \citet{panfilov2025strategic} show output-based red-team evaluations can be defeated by strategically dishonest model behaviour. Red-teaming generates evidence; it does not impose disclosure standards on the resulting claims.
& Pre-submission
& No, complementary \\[3pt]
Tiered reproducibility (this paper)
& Operates at review time at NeurIPS. Forces explicit declaration of what evidence supports each frontier safety claim, at what tier of disclosure, with mandatory scope statements. Does not produce new evidence but ensures published claims are scrutinisable to the limit allowed by their sensitivity.
& Review-time
& -- \\
\bottomrule
\end{tabularx}
\end{table}

\paragraph{Why these mechanisms compose, not compete}
Benchmarks generate evidence; red teams stress-test it; auditors verify
it externally; tiered reproducibility ensures that when any of this
evidence is presented at NeurIPS in support of a safety claim, the
disclosure required matches the strength of the claim. The four
mechanisms occupy different points in the evidence lifecycle. The
present proposal is the only one that operates at review time, and is
therefore the only one capable of changing what counts as
peer-reviewed-published-at-NeurIPS for this claim category.

\paragraph{Why tiered reproducibility is the right tool here}
The 2025--2026 evidence (Section~\ref{sec:problem}) shows that the bottleneck
is not absence of evaluation methods but absence of disclosure norms
that match claim strength. A frontier lab can run a sophisticated
red-team exercise, commission an external audit, and then publish a
paper at NeurIPS reporting only summary statistics. Standardised
benchmarks, third-party audits, and red-teaming all have valuable
roles, but none of them changes what authors are required to disclose
when they ask NeurIPS to confer the legitimacy of peer-reviewed
publication on a frontier safety claim. The tiered framework does.

\paragraph{Limitations of the comparison}
The four mechanisms in Table~\ref{tab:alternatives} are not exhaustive.
Other proposals (secure enclaves for evaluation
\citep{anderljung2023}, structured access programmes
\citep{black-box-2024}, mutual privacy frameworks
\citep{bucknall2025mutual}) are compatible with tiered
reproducibility and would strengthen the T2 mechanism in particular.
The present paper's contribution is review-time disclosure norms; the
broader infrastructure for safe sharing of sensitive artefacts is
genuinely complementary work.

\clearpage
\section{Federated Review Infrastructure}
\label{app:federated}
% Appendix F: Federated review infrastructure (full elaboration)

The Tier-2 confidentiality requirement does not depend on any single
institution. We propose a \textbf{federated colloquium of
secure-review entities}, including national AI Safety Institutes
(UK AISI, US AISI, Japan AISI, and analogous bodies emerging from the
EU AI Office and the India AI Mission), independent evaluators (METR,
Centre for AI Safety), and trusted academic secure-review labs. In
any given review cycle, a T2 submission is assigned to the entity
with the most relevant domain expertise and available capacity, not
to a geographically fixed host. CBRN uplift claims would route to a
chem--bio-equipped AISI; autonomous cyber-offence claims to METR or
an equivalent specialist evaluator; systemic-risk modelling claims to
an academic partner with the relevant infrastructure.

\paragraph{F.1 Why federation, not centralisation}
The single-host alternative concentrates four distinct risks at one
institution: capacity bottlenecks during high-submission cycles;
political fragility tied to that host's funding, statutory status, or
change of administration; expertise gaps when claims fall outside the
host's domain; and a perception that NeurIPS is delegating scientific
review to a national regulator. Federation neutralises each. The
choice of host is a function of claim domain, not geography; the
NeurIPS Confidential Panel (Appendix~\ref{app:governance}) retains
oversight of tier classification and adequacy statements across hosts,
preserving review consistency without centralising infrastructure.

\paragraph{F.2 Four properties of the federated design}
The federated colloquium has four properties the single-host
alternative does not. \emph{Parallel capacity}: multiple hosts
collectively sustain higher T2 volumes than any one panel could,
without any single panel becoming the bottleneck.
\emph{Political robustness}: no single AISI's funding cycle,
statutory status, or change of administration becomes a single point
of failure for the framework. \emph{Domain matching}: specialised
claims are routed to specialised reviewers, raising the validity of
the adequacy statement the panel produces. \emph{Institutional
humility}: NeurIPS coordinates rather than regulates; existing
secure-review infrastructure carries the operational load it is
already designed to carry.

\paragraph{F.3 Consistency with existing precedents and the 2026 governance landscape}
Federation matches the structure of the precedents the paper draws
on: ACM artefact-review badges \citep{acm2020artifact} are issued
under different evaluation committees by sub-community; clinical-trial
review is conducted by national regulators that recognise each other's
findings under formal cooperation arrangements; coordinated
vulnerability disclosure operates through a network of CERTs rather
than a single global authority. The framing is also consistent with
the 2026 governance landscape \citep{seoul2024,delhi2026,iaseai2026},
which treats AI safety review as a multi-institutional,
cross-jurisdictional activity rather than a unilateral national
responsibility.

\paragraph{F.4 Allocation protocol}
On receiving a T2 submission, the senior PC consults the federated
colloquium's domain registry and assigns the submission to the host
whose declared domain expertise best matches the claim's primary
sensitivity (CBRN, cyber-offence, systemic-risk modelling, deception,
self-replication). If no single host has a clear match, two hosts
co-review with one designated as primary. The host's panel produces
the adequacy statement under its own internal procedures; the NeurIPS
Confidential Panel reviews the statement for consistency with the
tier-classification framework before publication. Hosts publish an
annual workload report; the senior PC re-balances assignments in the
following cycle if any host approaches capacity. Authors may petition
for reassignment based on demonstrated conflict of interest, prior
adversarial engagement, or logistical infeasibility (e.g., legal
constraints on data transfer to a specific jurisdiction); petitions
are filed within seven days of host notification, and the final
reassignment decision rests with the senior PC.

\paragraph{F.5 Failure modes specific to federation}
Federation introduces failure modes a single-host design does not.
\emph{Inter-host inconsistency}: different hosts could apply different
adequacy thresholds. Mitigation: the NeurIPS Confidential Panel
audits a stratified sample of adequacy statements per cycle and
publishes inter-host calibration metrics. \emph{Race-to-the-bottom}:
authors could shop for the most lenient host. Mitigation: host
assignment is made by the senior PC, not by authors.
\emph{Colloquium contraction}: if the colloquium shrinks to one host,
federation reverts to centralisation. Mitigation: the Year-3 review
of host participation (Section~\ref{sec:impl}) is the formal trigger for
expanding the colloquium or scaling back the framework if host supply
drops below a viability threshold.

\begin{figure}[ht]
\centering
\begin{tikzpicture}[
  font=\sffamily\footnotesize,
  hub/.style={
    rectangle, rounded corners=2pt, draw=neuripsnavy, line width=0.6pt,
    fill=lightnavy, minimum height=9mm, minimum width=92mm,
    align=center, inner sep=3pt
  },
  audit/.style={
    rectangle, rounded corners=2pt, draw=tier2, line width=0.7pt,
    fill=tier2!10, minimum height=9mm, minimum width=92mm,
    align=center, inner sep=3pt
  },
  output/.style={
    rectangle, rounded corners=2pt, draw=neuripsnavy, line width=0.6pt,
    fill=lightnavy, minimum height=9mm, minimum width=92mm,
    align=center, inner sep=3pt
  },
  host/.style={
    rectangle, rounded corners=2pt, draw=tier1, line width=0.6pt,
    fill=tier1!10, minimum height=12mm, minimum width=20mm,
    align=center, inner sep=3pt, font=\sffamily\scriptsize
  },
  flow/.style={-{Latex[length=2mm]}, line width=0.55pt, draw=neuripsnavy!80},
  audline/.style={dashed, -{Latex[length=2mm]}, line width=0.5pt, draw=tier2!90},
  edgelab/.style={font=\sffamily\scriptsize, text=neuripsnavy!85, inner sep=1pt},
  audlab/.style={font=\sffamily\scriptsize\itshape, text=tier2!90, inner sep=1pt},
]

% Top band: audit/oversight layer
\node[audit] (panel) at (0, 4.0) {NeurIPS Confidential Panel: cross-host calibration audit};

% Coordination band: senior PC
\node[hub] (pc) at (0, 2.0) {NeurIPS senior PC: domain-matched assignment of T2 submissions};

% Hosts band: four hosts evenly spaced -- labelled by review domain, not institution
\node[host] (h1) at (-3.45, 0)  {Host A\\\textit{chem-bio /}\\\textit{CBRN review}};
\node[host] (h2) at (-1.15, 0)  {Host B\\\textit{cyber-offence}\\\textit{review}};
\node[host] (h3) at ( 1.15, 0)  {Host C\\\textit{deception /}\\\textit{autonomy review}};
\node[host] (h4) at ( 3.45, 0)  {Host D\\\textit{systemic-risk}\\\textit{modelling}};

% Colloquium frame around hosts
\begin{pgfonlayer}{background}
  \node[draw=tier1!50, dashed, line width=0.4pt, rounded corners=3pt,
        fit=(h1)(h4), inner sep=4pt, inner ysep=4pt] (collo) {};
\end{pgfonlayer}
\node[edgelab, text=tier1!85, anchor=north] at ($(collo.south)+(0,-0.5pt)$) {Federated colloquium of independent secure-review hosts};

% Bottom band: public output
\node[output] (out) at (0, -2.4) {Public adequacy statement: disclosure label issued with paper};

% --- Vertical assignment arrows: PC down to each host ---
\draw[flow] (pc.south -| h1.north) -- (h1.north);
\draw[flow] (pc.south -| h2.north) -- (h2.north);
\draw[flow] (pc.south -| h3.north) -- (h3.north);
\draw[flow] (pc.south -| h4.north) -- (h4.north);
\node[edgelab] at (-4.7, 1.1) {\itshape assign};

% --- Vertical adequacy arrows: each host down to public output ---
\draw[flow] (h1.south) -- (out.north -| h1.south);
\draw[flow] (h2.south) -- (out.north -| h2.south);
\draw[flow] (h3.south) -- (out.north -| h3.south);
\draw[flow] (h4.south) -- (out.north -| h4.south);
\node[edgelab] at (-4.7, -1.3) {\itshape adequacy};

% --- Single dashed audit arrow from Panel down to PC (then PC routes audit to colloquium per text) ---
\draw[audline] (panel.south) -- node[audlab, right=2pt, pos=0.5] {calibrate} (pc.north);

\end{tikzpicture}
\caption{Architecture of the federated colloquium for Tier-2 review.
A T2 submission flows top-to-bottom through three bands. The senior
PC performs domain-matched assignment to one host in the colloquium;
the host conducts confidential review under its own secure-review
procedures and returns an adequacy statement, which becomes the
basis for the paper's public disclosure label. The NeurIPS
Confidential Panel (top) audits a stratified sample of adequacy
statements across hosts and feeds calibration back through the
senior PC, maintaining inter-host consistency. NeurIPS coordinates
and audits; it does not own the secure infrastructure each host
already operates. Hosts are labelled by review domain rather than
institution; candidate hosts (national AI Safety Institutes, METR,
CAIS, academic secure-review labs) are discussed in the surrounding
text. Domain mappings are illustrative; assignment is determined
per-submission by the senior PC.}
\label{fig:federated}
\end{figure}
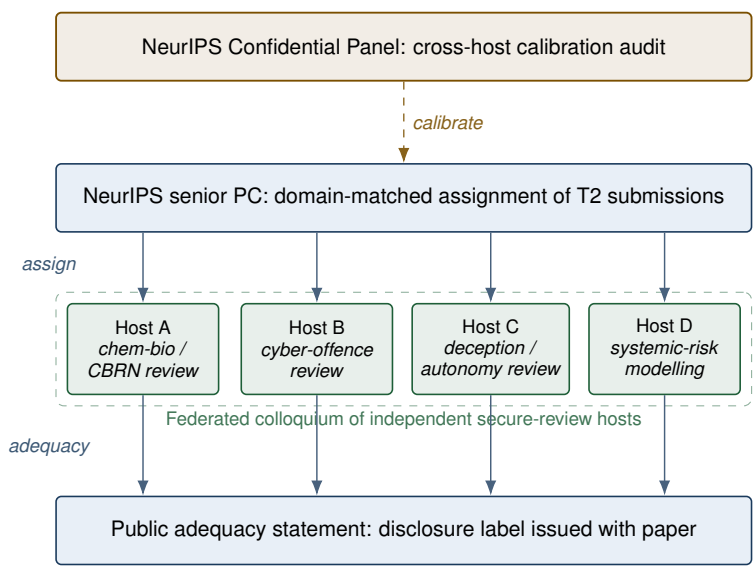

\paragraph{F.6 Architecture of the federated colloquium}
Figure~\ref{fig:federated} summarises the operational flow: a T2
submission enters through the NeurIPS senior PC, which performs
domain-matched assignment to one host in the colloquium; the host
conducts confidential review under its own internal procedures and
returns an adequacy statement; the NeurIPS Confidential Panel audits
the statement for consistency with the tier-classification framework
before publication. The colloquium itself is a flat network of
independent hosts; NeurIPS coordinates allocation and audits
adequacy, but does not own the secure infrastructure that each host
already operates.

\paragraph{F.7 The road ahead}
The federated colloquium is designed to extend, not just to operate.
The same architecture would absorb new claim categories as frontier
risks evolve (e.g., autonomous-agent oversight, post-deployment
monitoring of long-context misuse), and would degrade gracefully if
any single host withdraws: the senior PC re-routes to remaining
hosts with the relevant domain match, and the Year-3 review
(Section~\ref{sec:impl}) provides the formal trigger for expanding
the colloquium or scaling back the framework if host supply drops
below a viability threshold.

\end{document}